\documentclass[%
reprint,
superscriptaddress,
groupedaddress,
nofootinbib,
amsmath,amssymb,
aps,
prstab,
floatfix,
]{revtex4-2}

\usepackage{dblfloatfix}
\usepackage{graphicx}
\usepackage{dcolumn}
\usepackage{bm}
\usepackage{siunitx}
\usepackage{tablefootnote}
\usepackage{placeins}
\usepackage{xcolor}
\usepackage{appendix}
\usepackage{multirow}
\usepackage{hyperref}

\begin{document}

\preprint{APS/123-QED}

\title{Laser-Plasma Accelerator Beams in Light Sources: \\Femtosecond High-Brightness Radiation through Chirped Pulse Injection}
\author{S.~A.~Antipov\textsuperscript{1}\thanks{Sergey.Antipov@desy.de}}
\author{P.~Burghart\textsuperscript{1,2}}
\author{I.~Agapov\textsuperscript{1}}
\author{A. Mart{\'i}nez de la Ossa\textsuperscript{1}}
\author{W.~Leemans\textsuperscript{1,2}}

\affiliation{\textsuperscript{1}Deutsches Elektronen-Synchrotron DESY, Notkestr. 85, 22607 Hamburg, Germany}

\affiliation{\textsuperscript{2}University of Hamburg, Germany}

\date{\today}

\begin{abstract}

We propose a chirped-pulse injection scheme into a hard x-ray low-emittance synchrotron light source such as PETRA~IV from a laser-plasma electron injector with active energy compression. The scheme enables delivering kA-scale short pulses with several tens of hertz repetition rate to any synchrotron beamline in the ring and allows producing femtosecond temporally coherent radiation pulses at target beamlines. 


\end{abstract}

\maketitle


Synchrotron radiation sources are indispensable research instruments, whose brightness has been constantly improving over the past decades and has recently seen a quantum leap enabled by novel multi-bend achromat lattice concepts~\cite{Einfeld2014FirstMBA,Raimondi2023,Raimondi2021CommissioningHMBA,Leemann2009MAXIVBeamDynamics,Tanaka2016SPring8Upgrade,Streun2023SLSUpgrade,Borland2018APSUpgrade}. At the same time, their equilibrium bunch length and stored current cannot be significantly influenced by lattice design and remain largely unchanged, with typical values in the tens of ps range. Breaking this barrier and generating short (fs) radiation pulses in electron storage rings has attracted significant interest with various beam manipulation techniques such as laser slicing or longitudinal focusing with rf cavities having been proposed to accomplish the goal ~\cite{doi:10.1142/S1793626810000415}. These techniques are, however, usually limited to one beamline and are associated with significant additional hardware installations. They have been superseded by availability of high-brightness, low-emittance, and high-repetition-rate beams from electron linacs, such as those driving the state-of-the-art free-electron lasers (FELs)~\cite{Galayda2018LCLSII,Decking2020EuXFEL,Plath2016FLASHMultiplex}.

Beyond short-pulse operation, achieving longitudinal coherence in storage-ring sources has also faced  several difficulties. Producing steady-state coherent radiation has been so far limited to THz bursts~\cite{Evain2019SOLEILMicrobunchingControl,Brosi2019MicrobunchingLowCharge,Schreiber2026NegativeMomentumCompaction}, and FEL oscillators at largely optical wavelengths~\cite{Couprie1990SuperACOFEL,Hama1994UVSORFEL,Litvinenko1998DukeOK4FirstLasing,Walker2001ELETTRAFEL,Sei2011NIJIIVMIRFEL}. The development of coherent extreme ultraviolet (EUV) radiation through techniques such as steady-state microbunching has attracted significant attention~\cite{Kruschinski2024SSMBFoundation} but it remains in the development phase so far. It is limited to low-energy (MeV) machines, and faces significant technological challenges~\cite{Deng2021CourantSnyderLongitudinal,Deng2023BreakdownClassicalBunchLength,Deng2026SSMBTransverseLongitudinal}. These schemes are limited by several factors. First, the equilibrium bunch length and thus the achievable peak current limit the coherent wavelength reach. Second, the beam heating through coherent radiation should not exceed the synchrotron radiation damping, thus limiting the coherent radiation power. Last, imprinting steady-state density modulations, beyond requiring additional laser infrastructure, is only possible with dedicated low-momentum-compaction optics design, and is incompatible with state-of-the-art fourth-generation light source designs. Still, overcoming these limitations and producing short-pulse coherent radiation in storage rings remains an active area of research. Of particular interest is extending the modalities of existing large-scale facilities for most cost-efficient and versatile use of expensive x-ray beamline and accelerator infrastructure.

Laser-plasma acceleration (LPA) is now entering the application phase with recent successful demonstrations of both self-amplified spontaneous emission (SASE) and seeded FELs~\cite{Wang2021, Labat2023}. With advances in beam quality and phase-space manipulation, they are now capable of serving as injectors for future light sources~\cite{Agapov:2025irh}. Active energy compression allows the LPA beam to reach extremely low levels of energy spread and jitter, down to $\sim 10^{-4}$, by employing downstream radiofrequency (rf) cavities~\citep{Antipov:2021eko}. This scheme is envisioned for the PETRA~IV plasma injector~\cite{Agapov:2025irh} to enable clean and efficient injection in the storage ring. Synchrotron storage rings typically run in top-up mode and have many hours of beam lifetime. With $\sim 1$~\% requirements on beam current stability, the duty cycle of their injectors is in the few percent range, which creates an opportunity for additional applications of the high-brightness LPA injector.

In this paper we demonstrate a novel possible capability of the energy-compression-based LPA injector. Due to its energy spread, a short bunch would quickly spread out longitudinally when injected into a storage ring before reaching an experimental station due to a non-zero $R_{56}$ of the ring arc. The rf cavity of the injector energy compressor can, however, be adjusted to create a matched energy chirp and result in optimal bunch length at any target beamline in the ring. Such chirped pulse injection, in the example of the PETRA~IV ring, will deliver several kA peak current fs bunches in a single pass with the 30 Hz repetition rate, that could be cleared after one turn in the ring and be delivered simultaneously to the top-up operation. Beam dynamics simulations presented further show that taking into account realistic operation parameters such as energy jitter and ring momentum acceptance, as well as the coherent synchrotron radiation effect in the storage ring arcs, pose no problem to the scheme. The achievable level of compression is studied and it is shown that beyond short-pulse operation at any beamline without modification of the PETRA~IV facility. 
\newpage

\begin{figure*}[]
    \centering
    \includegraphics[width=0.99\linewidth]
    {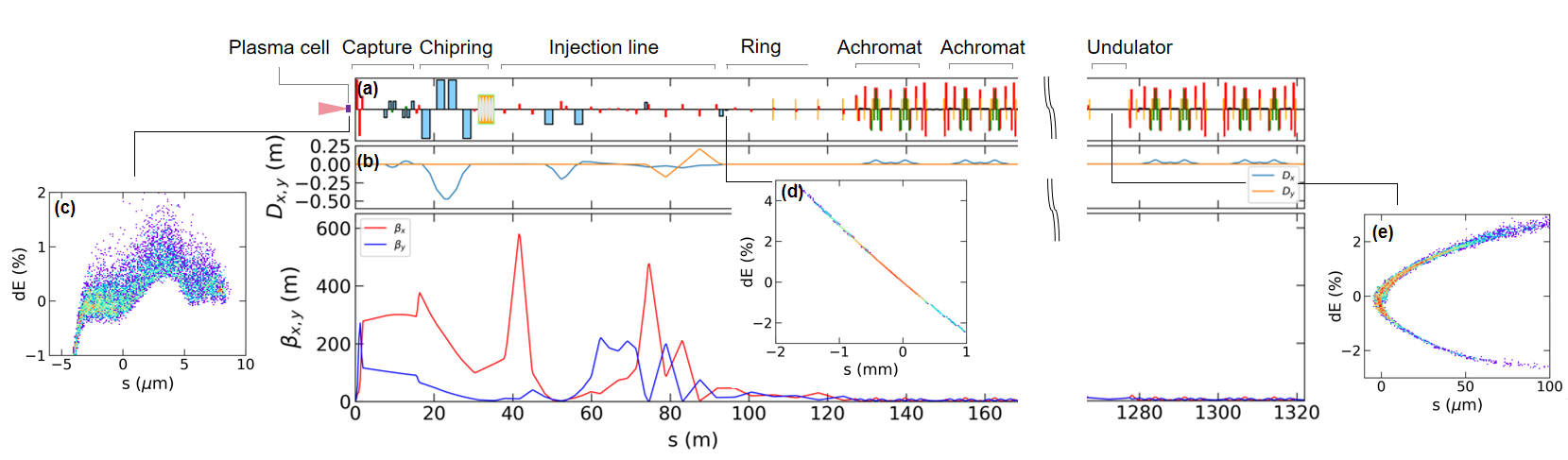}
    \caption{
    Schematic layout (a), optics functions (b) and start-to-end tracking simulations (c-e) of chirped LPA beam injection into the PETRA~IV light source. An electron bunch is created in an LPA with a short bunch length and a large energy spread (c). It is transported to the injection point through a beamline that manipulates the longitudinal phase space to impart a position-energy chirp (d). The bunch is then compressed in the ring achromatic cells before arriving at a target undulator beamline with a kA-scale peak current and fs-scale pulse length (e). Longitudinal beam distributions simulated in \texttt{Ocelot}. Plasma cell denoted in violet, focusing quadrupoles in red, bending magnets in blue, sextupole magnets in green, rf in orange.
}
    \label{fig:LPA_in_SSS}
\end{figure*}

Figure~\ref{fig:LPA_in_SSS} depicts the proposed LPA injector setup for 6~GeV PETRA ~IV, based on a design described in~Ref.~\citep{Agapov:2025irh}. It consists of an LPA followed by a quadrupole triplet to capture the electron beam from the plasma cell, a chicane with sextupoles for chromaticity correction, a large stretcher chicane, and an rf cavity for suppressing energy spread and correcting energy deviations. Finally, achromatic bends bring the beam into the storage ring tunnel where it is injected by a set of fast kickers and a pulsed septum.
During the nominal operation as an injector the rf voltage is set to cancel the energy chirp produced by the chicane and bring all particles to the design energy. This drastically reduces the relative energy spread of the beam from $\sim 10^{-2}$ to $< 10^{-4}$, allowing it to fit within the ring's energy acceptance and ensuring a high injection efficiency.

In nominal energy compression mode, the bunch reaches the septum already stretched to an rms length $\sigma_s \sim 1$~mm, with $\sigma_\delta < 10^{-4}$. Once captured, it undergoes synchrotron oscillations and ultimately damps toward the ca. 12 mm rms equilibrium bunch length. Conversely, a hypothetical uncompressed LPA bunch would reach the septum with $\sigma_{s,0} \simeq 3~\mu$m and $\sigma_{\delta,0} \simeq 10^{-2}$, but the $R_{56}$ of the first octant would lengthen it by about $100~\mu$m in one pass, strongly reducing its peak current before the bunch reaches a user beamline. We propose that the bunch is instead stretched in the injector and given a controlled negative energy-position chirp, so that the ring arcs initially compress it.

The rf voltage in the injector compressor can be adjusted to produce an arbitrary desired chirp $h = \partial \delta / \partial s$. Here we use $s$ as the longitudinal coordinate relative to the reference particle, positive toward the bunch tail, and $\delta = (E-E_0)/E_0$. Knowing the transfer map between the injection point and the target beamline, the injection energy chirp can be matched to produce maximum bunch compression at a chosen undulator:
\begin{equation}
    h = -1/{R_{56}} = -1/({R_{56}^{arc} + R_{56}^{tl})}, 
\end{equation}
where $R_{56}^{arc}$is the momentum compaction in the arc of the storage ring, defined as $R_{56} \equiv \partial s / \partial \delta$. An extra factor $R_{56}^{tl}$ represents the momentum compaction in the injection line after rf~\cite{ftnt:impact_of_injection_line}. The required rf voltage to generate this chirp is given by 
\begin{equation}
    h = 1/{R_{56}^{ch}} + k_{rf}Ue/E_0, 
\end{equation}
where $R_{56}^{ch}$ is the momentum compaction of the LPA injector decompression chicane and $k_{rf}$ and $U$ are the rf wave vector and voltage respectively.
$R_{56} > 0$ for the above-transition PETRA~IV ring, whereas $R_{56}^{ch}< 0$ for the decompression chicane. The chicane therefore generates a negative chirp, and the rf kick adjusts its magnitude to the value required at the target beamline. As the $R_{56}^{arc}$ is relatively small, the rf has to increase the energy spread of the LPA beam to enable longitudinal compression in the ring. Consequently, achieving maximum compression at nearer beamlines in the ring requires a%

\begin{table}[!h]
    \centering
    \caption{
    Beam parameters at different stages of tracking. Target beamline U61.} 
    \begin{tabular}{lrrr}
        \hline\hline
         Parameter & LPA exit & Injection & Beamline \\
         \hline
         Bunch charge & 87~pC & 78~pC & 75~pC\\
         Peak current & 3.7~kA & 31~A & 2.2~kA\\
         Bunch length (core), rms & 3.14 $\mathrm{\mu m}$ & 0.44~mm & 2.3 $\mathrm{\mu m}$\\
         Energy chirp & 0 & 26.5 $\mathrm{m^{-1}}$ & $\infty$ \\
         Rel. energy spread, rms & 0.46~\% & 1.2\% & 1.1\%\\
         Hor. norm. emit., rms & 4.15 $\mathrm{\mu m}$ & 3 $\mathrm{\mu m}$ & 3.5 $\mathrm{\mu m}$\\
         Vert. norm. emit., rms & 1.66 $\mathrm{\mu m}$ & 1.5 $\mathrm{\mu m}$ & 3.5 $\mathrm{\mu m}$ \\
         \hline \hline
    \end{tabular}
    \label{tab:beam_par}
\end{table}
\newpage
\onecolumngrid
\newpage
\twocolumngrid
\newpage
\onecolumngrid
\begin{figure*}
    \centering
    \includegraphics[width=0.99\linewidth]{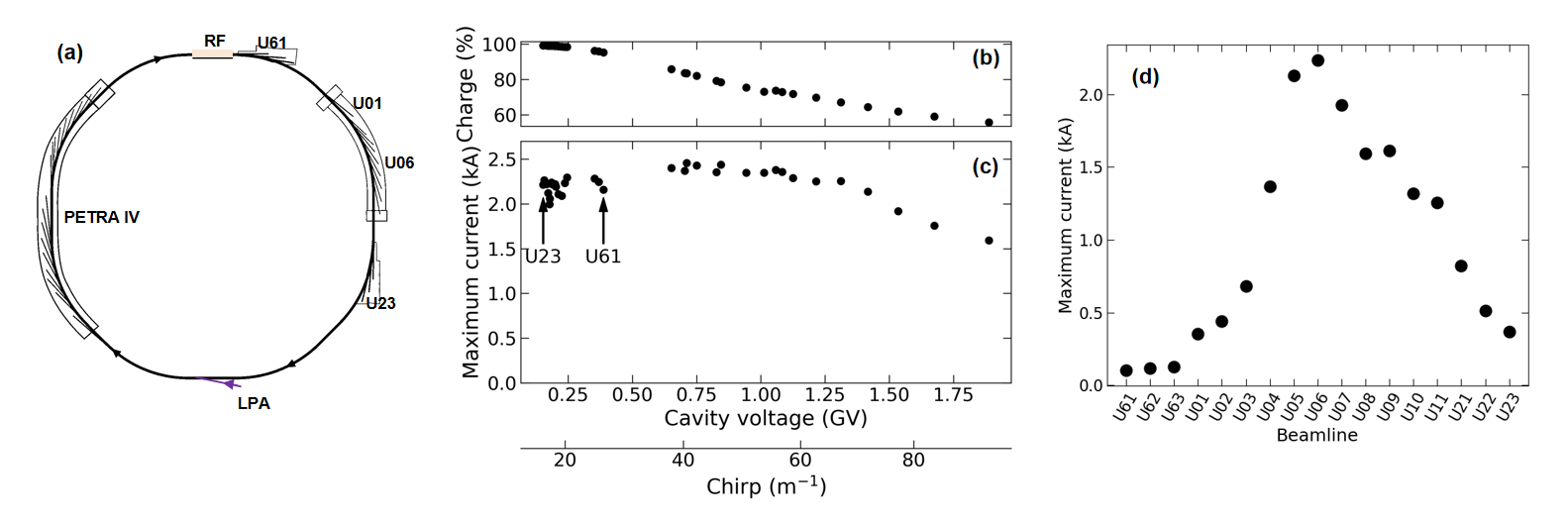}
    \caption{Schematic of the PETRA~IV light source showing the locations of its beamlines and the LPA injector (a). Peak current and charge transmission achievable at beamlines around PETRA~IV and the corresponding charge transmission into the storage ring vs X-band rf chirping voltage (b,c). Greater peak currents are achievable at Eastern beamlines U61--U23 after nearly a full revolution in the storage ring. Kiloampere-scale peak currents can be generated at several neighboring beamlines at the same time (d). Peak current dependence on beamline location for a single chirp value optimized to produce the largest peak current at the beamline U06: rf voltage 199.5~MV, chirp $18.36$~m$^{-1}$.}
    \label{fig:beamlines}
\end{figure*}
\newpage
\twocolumngrid

\noindent larger rf voltage than at those located further away. At $U = -E_0/ek_{rf}R_{56}^{ch}$ one obtains $h = 0$ and the condition for optimal energy compression of~Ref.~\cite{Antipov:2021eko}. At $h \neq 0$ the final projected energy spread is
\begin{equation}
    \sigma_\delta \approx \sigma_{\delta,0} |R_{56}^{ch}k_{rf}Ue/E_0 + 1| = \sigma_{\delta,0} |h R_{56}^{ch}|.
\end{equation}
For an efficient, lossless transport in the ring $\sigma_\delta$ shall be significantly smaller than its energy acceptance, typically a few percent. 

Using the same rf system that is used to improve injection efficiency during the nominal injector operation for imparting the chirp to the beam offers an elegant, though not the only way to extend the injector capability. Alternative approaches are briefly discussed in App.~\ref{app:Alt}.

The injection process was modeled starting from the plasma source through the LPA injector beamline and through one turn of the storage ring. A realistic particle distribution was obtained through Bayesian optimization of the LPA stage using the particle-in-cell code \texttt{FBPIC}~\cite{bib:fbpic} and the optimization software \texttt{Optimas}~\cite{FerranPousa2023,optimas}. It has 87~pC of charge, an rms bunch length of 3.14~$\mu$m, and $10^{-2}$ rms energy spread (Table~\ref{tab:beam_par}). The beam is then tracked through the LPA beamline and ring arcs using the \texttt{Ocelot}~\cite{Agapov2014OCELOT, Tomin2017OCELOT} macroparticle tracking code. 

Optical functions of the LPA beamline matched to the arc of the main ring are shown in Fig.~\ref{fig:LPA_in_SSS}b. The tracking was carried out with $10^4$ macroparticles and second-order transfer maps, which are sufficient for single-pass simulations. An aperture model of the storage ring and its injection elements was included, as well as coherent synchrotron radiation. Wakefields, incoherent synchrotron radiation effects, and ring rf were neglected; their impact is negligible as discussed later. The maximum achievable peak current per PETRA~IV beamline is shown in Fig.~\ref{fig:beamlines}c. 
As expected, the beamlines further away from the injection point can achieve higher peak current with lower rf voltage and better transmission, and are more favorable for this scheme than the beamlines directly downstream of the injection point. The first downstream beamlines require large chirp values, where the finite ring's momentum acceptance limits the transmission (Fig.~~\ref{fig:beamlines}b). This is not a fundamental limitation, but determined by the physical apertures, mainly that of the injection septum.

\begin{figure}[h]
    \centering
    \includegraphics[width=0.99\linewidth]{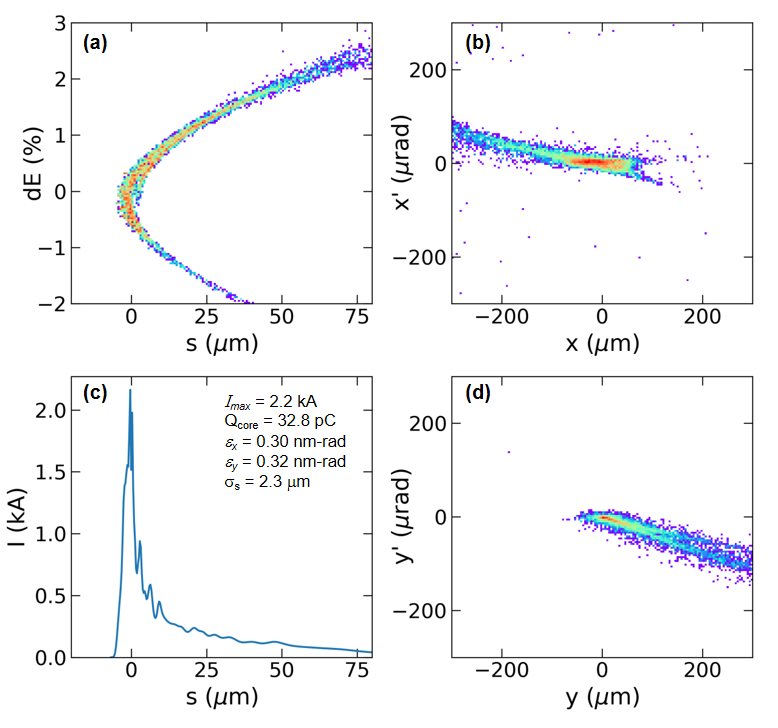}
    \caption{An example phase space of the ideally compressed bunch core at beamline U61: the longitudinal (a) the two transverse planes (b,d). The peak current reaches 2.2 kA in the bunch core (c).}
    \label{fig:phase_space_U23}
\end{figure}

The bunch phase space at one of the beamlines is shown in Fig.~\ref{fig:phase_space_U23}. The beam distribution fully fits within the ring acceptance. The longitudinal distribution exhibits a non-Gaussian structure typical of rf compression, with a low-emittance high-current core. The high-density bunch core has an rms length of 2.3~$\mu$m and 33~pC of charge, reaching 2.2~kA peak current. Its horizontal and vertical emittances are 0.30 and 0.32~nm  (App.~\ref{app:num_meths}).

While the chirp value can be optimized for a particular beamline, large peak currents would still be achievable at many other beamline locations. Figure \ref{fig:beamlines}d shows peak currents at several beamlines with a setting optimized for a particular location. High peak currents in excess of 1 kA can be achieved at several beamlines simultaneously thanks to the relatively small $R_{56}$ between the neighboring beamlines.

Several effects such as LPA energy jitter, synchrotron radiation, and beam acceleration by the ring's rf system will affect the chirp and compression of beams in reality, beyond the simplified picture described above. They will, however, have only a moderate impact on the peak current for the parameters of PETRA~IV and its LPA injector. For more details, see App.~\ref{app:jitter}--\ref{app:rf}.

\begin{figure}[h]
    \centering
    \includegraphics[width=0.99\linewidth]{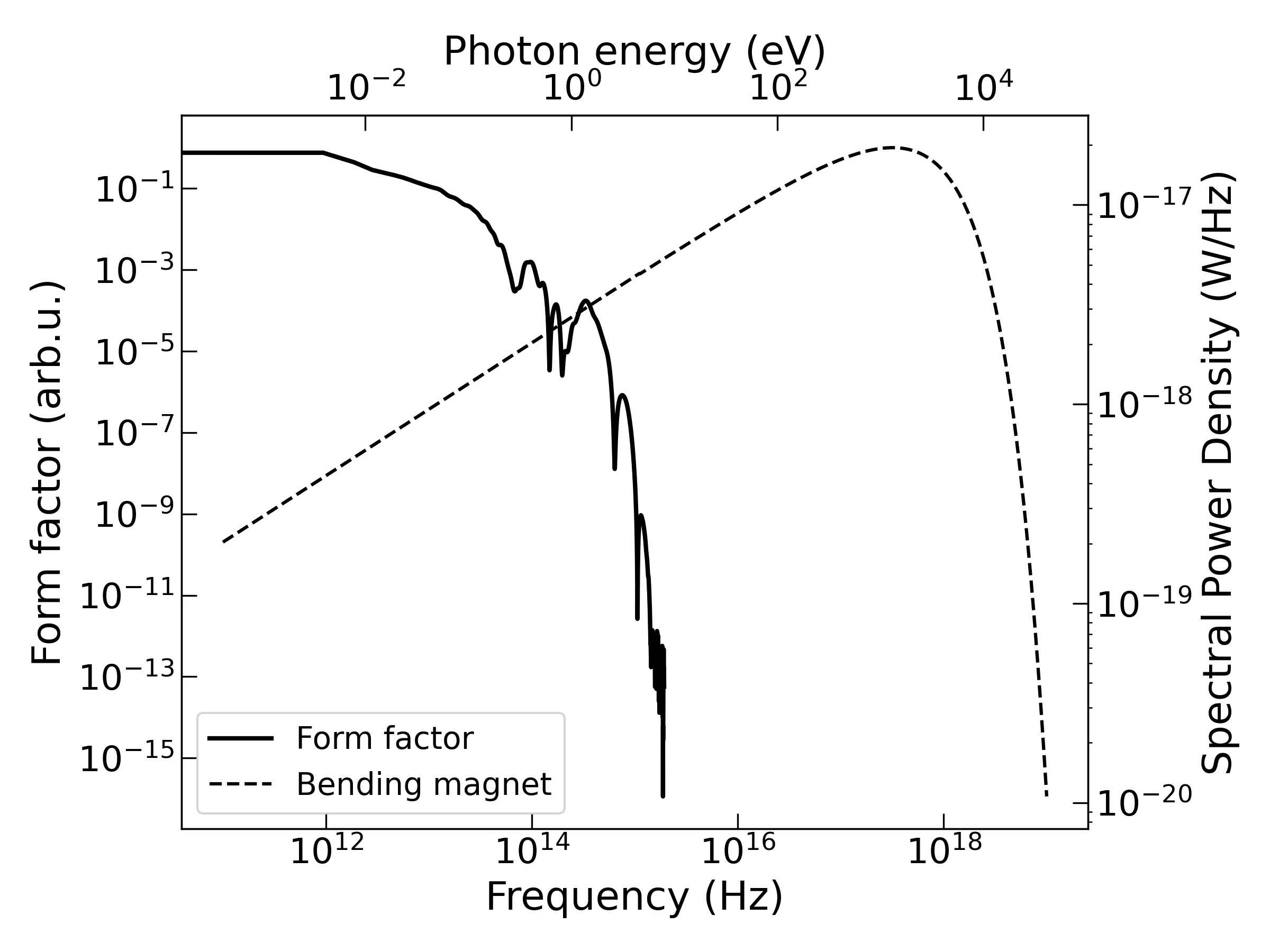}
    \caption{An example of expected radiation form factor and bending magnet spectrum from a 0.2 T bending magnet at the location of beamline U61.}
    \label{fig:spectrum_U61}
\end{figure}

The compressed bunches will produce unique photon radiation in the storage ring. The coherent spectrum extends from the THz range into the near-UV, reaching a few $10^{15}$~Hz, or about 10 eV~(Fig.~\ref{fig:spectrum_U61}).
This allows harvesting coherent THz to optical to near UV radiation with a simple bending magnet. The pulse duration is defined by the compressed bunch length of only about 8~fs rms.

Interestingly, the high peak current offered by the compressed LPA bunch offers hope for making a single-pass FEL process in the ring feasible. Yet, it might be challenging. Consider again the PETRA~IV example: a suitable location for the FEL undulator in the PETRA~IV ring is its flagship beamline U61. Our simulations predict that the LPA beam can be transported to this location nearly without losses and reaching around 2.2~kA peak current at the undulator (Fig.~\ref{fig:phase_space_U23}c). The relative energy spread in the bunch core is just $\sigma_\delta = 3.6\times 10^{-3}$, consistent with the $\sim 3\times 10^{-3}$ estimate of Eq.~(3). One appealing target wavelength for an FEL would be 13.6~nm in EUV. Parameters of a suitable EUV undulator device are listed in 

Table~\ref{tab:undulator}. For this device one obtains a Pierce parameter $\rho = 4.8\times10^{-3}$, at the border of the $\sigma_\delta \lesssim \rho$ condition, necessary for the FEL process. A 3D estimate of gain length according to Ming-Xie model~\cite{NguyenMarksteiner2014OneDimensionalFELTheory} yields 3~m, allowing a moderate coherent amplification over the length of the undulator.

The \texttt{Genesis} setup consisted of one 10-m-long undulator with a betatron waist in the middle. No significant coherent amplification was observed over the 10~m length of the undulator device~\cite{ftnt:super_ID_length}, which can be explained by the need for $\sim 4-5$ gain lengths for the self-amplified spontaneous emission (SASE) process to settle. The FEL process could be aided by longer strings of undulators and further improvement of LPA beam quality (reduction of emittance, slice energy spread). 
A 4–5-orders-of-magnitude exponential increase in pulse energy was observed over a 25-m-long undulator when the normalized emittance was reduced by half, to $1.5~\mu$m. An increase of LPA bunch charge is unlikely to amplify SASE FEL gain significantly as the final peak current at the undulator is limited by the CSR in the preceding ring cells (App.~\ref{app:CSR}).

After producing coherent synchrotron radiation at the target beamline(s) the LPA bunch 
can be immediately dumped so that its signal does not interfere with measurements at other beamlines. At PETRA~IV this clearing can be done with its fast injection kickers capable of targeting individual 2~ns rf buckets. The coherent emission process can then be repeated at the frequency of the LPA injector, i.e. $10-30$~Hz (the exact rate depends on the choice of the laser system), as the storage ring needs to be refilled only once every few minutes to maintain its stored beam current. Note that the tens of Hz repetition rate is comparable to that of warm linacs.

\begin{table}[]
    \centering
    \caption{Possible super-ID undulator parameters considered for the FEL estimates.}
    \begin{tabular}{l l r}
    \hline
    \hline
         Parameter  & Symbol        & Value \\
    \hline
         Period     & $\lambda_u$   & 5.96~cm\\
         Field      & $B$           & 2.0~T\\
         Wavelength & $\lambda$     & 13.6~nm\\
         Length     & $L_u$         & $20-25$~m\\
         Strength   & $K$           & 11.1\\
         Pierce par.& $\rho$        & $4.8\times10^{-3}$\\
         Gain len. 3D & $L_{FEL}^{3D}$ & 3 m\\
    \hline
    \end{tabular}
    \label{tab:undulator}
\end{table}

The radiation generated by the LPA bunch can be used together, in synergy with the radiation of the stored beam. The most promising scenarios seem to be various timing modes, where the ring operates with a small number of high-charge stored bunches. The baseline timing mode at PETRA~IV has 80 equidistant 8~nC bunches with 96~ns bunch spacing and other modes with a smaller number of bunches, larger bunch charge and longer bunch spacing are under consideration. By tuning the injection timing one can generate an arbitrary separation from $\sim 2$~ns to $\sim 96$~ns between the short, low-charge, high-peak-current LPA bunch and a long, high-charge bunch of the timing mode.

\textit{In conclusion,} chirped-pulse injection allows extending the capabilities of a plasma injector beyond its primary purpose of filling the light source ring and topping up its charge. The method takes advantage of longitudinal phase space manipulation to re-create the high, kA-scale peak currents offered by the LPA beams at the user beamlines in the storage ring. These high peak current femtosecond bunches can be delivered at a repetition rate as high as 30~Hz, which is one of the options for the LPA injector drive laser, `for free' to photon science users at any synchrotron beamline in the ring and letting no one be forgotten.  
The plasma injector thus becomes a versatile tool opening extra capabilities at the light source.

In contrast to present and past approaches to exploit synergies between injectors, FELs, and storage rings, such as the SPPS facility at SLAC~\cite{BENTSON2003205} and the SPring-8/SACLA facility~\cite{Yabashi2015SACLA}, our method will deliver high current, short pulses directly to photon beamlines in the storage ring, thus allowing use of an already existing experimental infrastructure.

Another strength of chirped pulse LPA injection lies in the possibility of exploiting synergies with high-charge timing mode filling patterns of the storage ring. By tuning the injection timing one can create an arbitrary spacing in the $\sim 1-100$~ns range between the LPA bunch and a $\sim 10$~nC bunch of the timing mode. This provides an opportunity for timing experiments with the stored beam. Promising scientific cases are those where an ultrafast pump prepares a well-defined initial state and then the key observables evolve naturally on nanosecond scales such that a hard X-ray probe at $1-100$~ns provides decisive structural information. 

Finally, while SASE lasing at EUV wavelengths about 13.5~nm seems beyond limits of the present technology, with further advances in bunch quality and stability on one side and in short-period high-field undulators on the other side one may eventually achieve lasing at dedicated beamlines in the storage ring. 
The repetition rate of such an FEL would ultimately be limited by the LPA drive laser; at 30 Hz, it would be competitive with conventional warm linacs.

\bibliography{biblio} 

\appendix
\section{Alternative schemes to produce chirped-pulse injection}\label{app:Alt}

A similar gymnastics in the longitudinal phase space can be performed with a positive $R_{56}$ chicane. 
This approach uses the large natural energy spread of the LPA beam to rotate its longitudinal phase-space distribution. 
It has several disadvantages though. Firstly, it would not allow correcting the central energy jitter of the LPA beam, which could be as large as 1\% or greater. In contrast, the active energy compression reduces central energy deviations significantly. Secondly, the positive $R_{56}$ chicane produces significant second-order $R_{56}$ and chromatic aberrations; these effects are detrimental to performance and their correction is not trivial as pointed out in \cite{tanaka:ipac2025-mopb033}. A detailed study of an $R_{56}^{ch} > 0$ alternative goes beyond the scope of this work.

Energy compression, as well as chirping described here can also be done elegantly with compact active plasma cells~\cite{FerranPousa2022}. Plasma lenses are capable of imparting very large energy chirps on the beam. On the other hand, due to the small decompression ($R_{56}^{ch} \sim -1$~mm), this method produces very short, ca. $10~\mu$m bunches. In practice this means that this method is best suited to feed undulator beamlines that are directly downstream of the injection point. In the PETRA~IV case the closest beamline is $R_{56} \approx 8$~cm away, which requires an energy chirp of $|h| \approx 100~\mathrm{GeV/m}=10^{-4}~\mathrm{GeV/\mu m}$. Given the short bunch length, the correlated relative energy variation across the bunch would be of order $10^{-4}$, smaller than the rms relative energy spread of order $10^{-3}$.

\section{Determination of beam parameters from numerical data}\label{app:num_meths}
To obtain an accurate measure of peak bunch current, the bunch current profile was determined using a Gaussian kernel density estimation, with the Gaussian kernel having a width of $\sqrt{N}$ particles with $N$ being the total number of particles in the bunch. In our case of beams with a very narrow core and long tails, this method is superior to bins of fixed width which would not capture the beam core in sufficient detail for an accurate peak current determination. Emittance was calculated statistically within a $1\sigma_s$ interval surrounding the position of peak current. Rare outliers with abnormally high values in $x',y'$, which might have affected the calculation of rms emittance, were discarded using a $4\sigma_{x',y'}$ cutoff.

\section{Impact of LPA energy jitter}\label{app:jitter}

\begin{figure}[h!]
    \centering
    \includegraphics[width=0.99\linewidth]{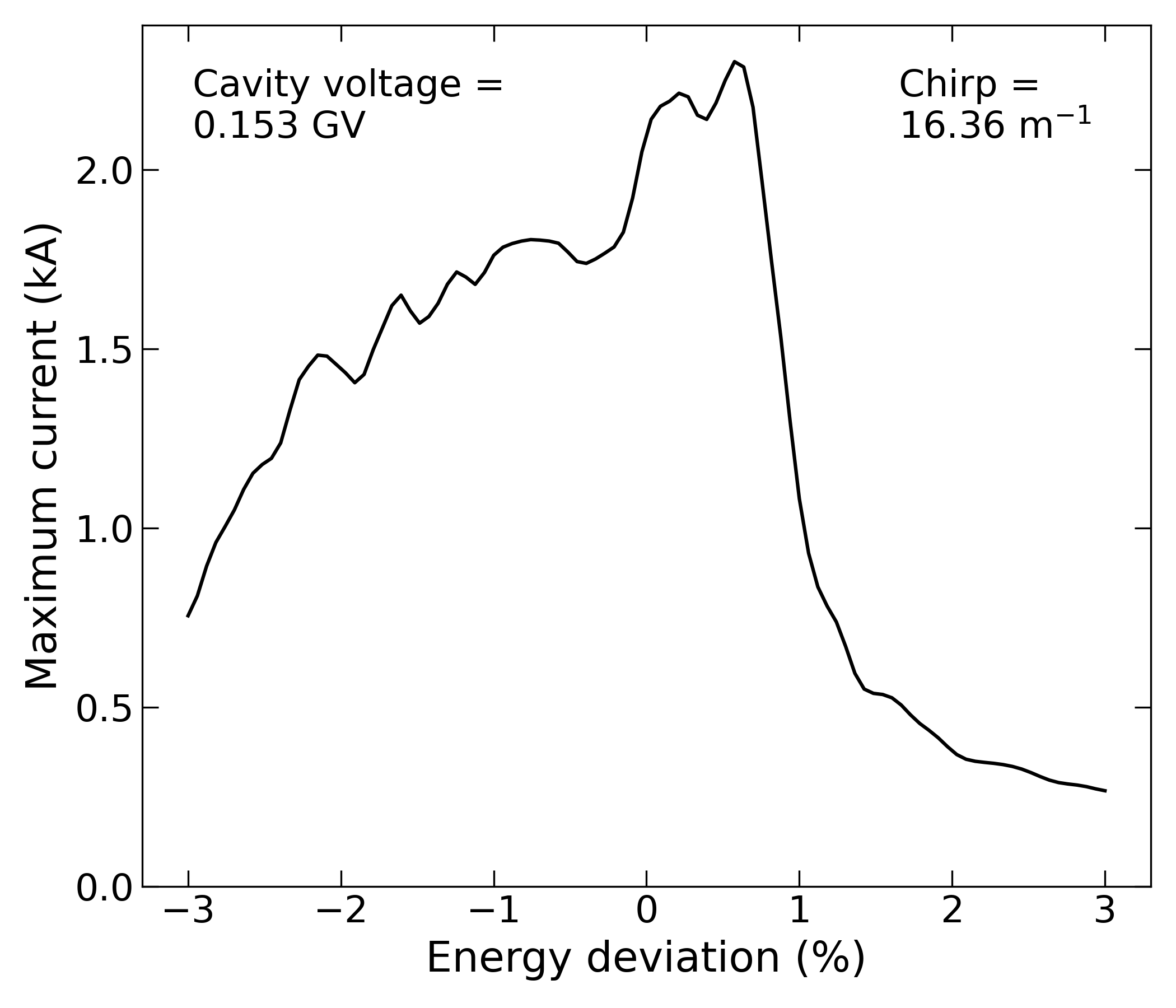}
    \caption{Tolerance of the maximum current observed at the PETRA IV beamline U23 to initial energy jitter at the LPA.}
    \label{fig:LPA_energy_jitter}
\end{figure}

The energy chirp of the LPA beam after the rf kick is affected by jitters in laser-to-rf timing, rf voltage and phase, and by the LPA energy jitter~\cite{Antipov:2021eko}. The  dominant of these effects is the LPA energy jitter that shifts the phase at the rf cavity and thus the chirp due to the nonlinearity of the rf kick. The change in phase $\delta\phi = k_{rf} R_{56}^{ch} \delta$ is approximately 14 deg for a 1\% energy offset and cannot be neglected. While the performance of the 6 GeV plasma injector is still to be quantified, the tolerance studies of peak current at the beamline as a function of energy deviation in the plasma cell indicate that the expected sub-percent stability of the plasma source will have a mild impact on the peak current, as shown in Fig.~\ref{fig:LPA_energy_jitter}. To produce this figure we artificially changed the initial LPA beam energy while leaving all other parameters unchanged in our simulation setup.

\section{Impact of coherent synchrotron radiation}\label{app:CSR}

For high-energy high-peak-current beams, traveling on curved trajectories, the coherent synchrotron radiation (CSR) is the most prominent potential source of beam quality degradation. 
To study the impact of CSR we performed tracking simulations with and without CSR in  \texttt{Ocelot} and \texttt{Elegant}~\cite{osti_761286}, with simplified Gaussian beams in \texttt{Elegant} and realistic LPA distributions in \texttt{Ocelot}.

In the \texttt{Elegant} simulations we progressively increased the LPA bunch charge while targeting the first beamline downstream of the injection point. We used simplified Gaussian beams since a realistic LPA beam distribution would vary with its charge. The initial distribution had $10^5$ particles. The results indicate no significant impact on energy spread, bunch length, or emittance up to 1~nC per bunch, more than an order of magnitude beyond the baseline LPA injector parameters. The rms energy spread and bunch length remained the same for 10 and 100~pC cases and increased by only about 10\% for the 1~nC case.

\begin{figure}[]
     \centering
     \includegraphics[width=0.99\linewidth]{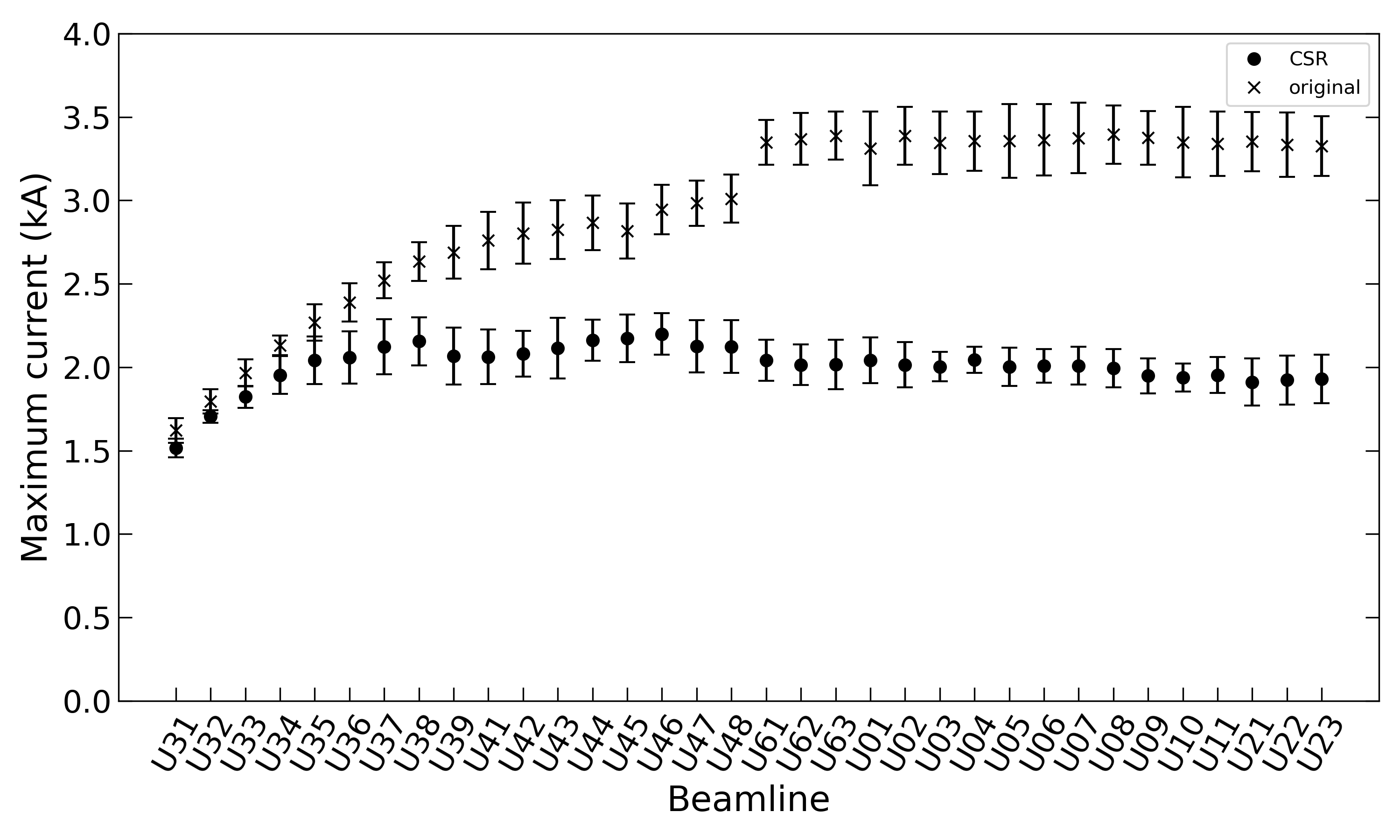}
     \caption{Coherent synchrotron radiation reduces the maximum achievable peak current compared to the no-CSR case. The effect becomes greater the further the beamline is from the injection point. Numerical simulation in \texttt{Ocelot}. For each undulator beamline we scanned the rf voltage around the optimal setpoint and report the mean values (dots and crosses) and rms spreads (error bars).}
     \label{fig:SCR_E_t}
 \end{figure}

In the \texttt{Ocelot} simulation setup we investigated true LPA bunch distributions of a fixed 87~pC charge while targeting sequentially every beamline in the ring, with and without CSR. In agreement with our \texttt{Elegant} results CSR produced no significant impact on the peak bunch current for the first few beamlines, after which the maximum current achievable without CSR started consistently exceeding that with the effect included (Fig.~\ref{fig:SCR_E_t}). While without the CSR effect we were able to achieve the peak currents of $\sim 3.5$~kA, comparable to that of the current at the plasma cell, at the most downstream beamlines, with CSR the peak current saturated around a smaller value of $2$~kA.

The 2~kA CSR limit is consistent with our order-of-magnitude estimates. Per unit length, the effect can be estimated through the CSR wake function ~\cite{bib:CSR}:
\begin{equation}
    W_{\rm CSR} = N_b r_e m_e c^2 \kappa^{2/3}\sigma_s^{-4/3},
    \label{eq:CSR}
\end{equation}
where $N_b$ is the bunch population, $r_e$ -- classical electron radius, $m_e$ -- electron mass, and $\kappa$ -- the curvature of the bending field. For the parameters of the optimally compressed beam with CSR (2.2~kA, $\sim 2.3~\mu$m rms bunch length) $W_{\rm CSR} \sim 0.5$ MV/m. This value is insignificant compared to the initial energy chirp of the order of $10^4$~MV/m. Thus, one shall not expect a meaningful impact of CSR when the beamlines closest to the injection point. When targeting the beamlines far downstream of the injection point though the initial chirp is necessarily smaller (Fig.~\ref{fig:beamlines}c), and at the same time the bunch is traveling through a larger number of bends. The CSR wake produces an additional energy deviation of $\Delta\delta \sim 10^{-4}/$m. Given about 6~m of bends per cell and the cell $R_{56}^{cell} \approx 1$~mm, it results in a change of bunch length of $\sim 0.6~\mu$m, which is comparable to the rms bunch length. Therefore, one may expect the combined CSR of several cells to have an effect on the peak current, as observed in simulation.

The CSR effect would become more prominent if the scheme is applied to low-energy rings with significantly larger curvature, as seen from Eq.~(\ref{eq:CSR}). This puts PETRA~IV and other high-energy rings in a unique position to serve such short high-current bunches.

\section{Incoherent synchrotron radiation and main rf}\label{app:rf}
As the bunch orbits the ring it loses energy due to synchrotron radiation. As long as the process is incoherent it does not affect the energy chirp but only the mean energy of the bunch. In PETRA~IV the bunch loses 4~MeV per turn, or about $6.7\times10^{-4}$ of its energy. If deemed necessary, this energy change can be compensated by a small adjustment of the injection beam energy.

Once per revolution this lost energy is compensated by the ring's rf system. In PETRA~IV the rf system is located in the North straight section, directly opposite to the LPA injection point (Fig.~\ref{fig:beamlines}a). It operates at 500~MHz and a nominal voltage of 8~MV. The rf focusing of the cavities affects the energy chirp but this effect is negligible compared to the initial chirp. The rf system produces a chirp $\sim 1$~MV/m, while the initial chirp is four orders of magnitude greater, $\sim 10^4$~MV/m.

\end{document}